\documentclass[aps,twocolumn,groupedaddress,superscriptaddress,amsmath,amssymb,prx]{revtex4-2}
\usepackage{amsmath}
\usepackage{graphicx}
\usepackage{dcolumn}
\usepackage{bm}
\usepackage{bbold}
\usepackage[usenames,dvipsnames]{color}
\usepackage[colorlinks,citecolor=Blue,linkcolor=Red, urlcolor=Blue]{hyperref}
\usepackage{tikz}
\usetikzlibrary{decorations.pathmorphing}
\usetikzlibrary{arrows.meta}
\usepackage{braket}
\usepackage[normalem]{ulem}
\usepackage{yfonts,dsfont,bbm,eufrak}

\newcommand{\be}{\begin{equation}}
\newcommand{\ee}{\end{equation}}
\newcommand{\ba}{\begin{array}}
\newcommand{\ea}{\end{array}}
\newcommand{\bqa}{\begin{eqnarray}}
\newcommand{\eqa}{\end{eqnarray}}

\usepackage{eucal}
\newcommand{\he}{\mathcal{H}}
\newcommand{\hh}{h}

\usepackage{eucal}
\newcommand{\Hf}{\mathcal{H}}

\begin{document}

\title{Quantum Simulation of Three-Body Interactions in Weakly Driven Quantum Systems}

\author{Francesco Petiziol}
\email{francesco.petiziol@unipr.it}
\affiliation{Universit\`a di Parma, Dipartimento di Scienze Matematiche, Fisiche e Informatiche, I-43124 Parma, Italy}    
\affiliation{UdR Parma, INSTM, I-43124 Parma, Italy}

\author{Mahdi Sameti}
\affiliation{Blackett Laboratory, Imperial College London, London SW7 2AZ, United Kingdom}

\author{Stefano Carretta}
\affiliation{Universit\`a di Parma, Dipartimento di Scienze Matematiche, Fisiche e Informatiche, I-43124 Parma, Italy}       
\affiliation{UdR Parma, INSTM, I-43124 Parma, Italy}

\author{Sandro Wimberger}
\affiliation{Universit\`a di Parma, Dipartimento di Scienze Matematiche, Fisiche e Informatiche, I-43124 Parma, Italy}       
\affiliation{INFN, Sezione di Milano Bicocca, Gruppo Collegato di Parma, Parma, Italy}

\author{Florian Mintert}
\affiliation{Blackett Laboratory, Imperial College London, London SW7 2AZ, United Kingdom}


\begin{abstract}
The realization of effective Hamiltonians featuring many-body interactions beyond pairwise coupling would enable the quantum simulation of central models underpinning topological physics and quantum computation.
We overcome crucial limitations of perturbative Floquet engineering and discuss the highly accurate realization of a purely three-body Hamiltonian in superconducting circuits and molecular nanomagnets.
\end{abstract}                              
\maketitle

\twocolumngrid 

\indent 
The behavior of quantum systems can be changed profoundly when subject to external driving \cite{Grifoni1998,Rabitz2010,Glaser2015,Odelin2019}. This is routinely utilized for quantum simulations \cite{Eckardt2017,Georgescu2014}, that is, to make a system emulate a different and more complex one, or to effectively eliminate the interaction between system and environment \cite{Viola1999,Lidar2014}. Of particular appeal for quantum simulations or quantum technological applications is the possibility to realize effective processes that do not exist in the undriven system.

A central goal in quantum simulations is the realization of interactions beyond the pairwise interactions of low-energy physics.
Driven systems can be characterized by effective Hamiltonians that include three-body or higher-order interactions. Hamiltonians of such types would enable, for instance, the quantum simulation of models of high-energy physics \cite{Zoller2019,Schweizer2019,Zoller2013,Zohar2017,Pedersen2020}, the implementation of quantum error correction algorithms \cite{Terhal2015,Kitaev2003,Dai2017}, or the realization of topologically protected states \cite{Jotzu2014,Rudner2019,Eckardt2017,Potirniche2017}.

Typical strategies to engineer these models effectively are based on periodic forcing and the use of high-frequency expansions \cite{Goldman2014,Bukov2015,Eckardt2015,Mikami2016}, which are appealing for many reasons. For instance, since three-body interactions result from two consecutive two-body processes, they appear as low-order contributions in such perturbative expansions \cite{Lee2016,Potirniche2017,Claassen2017, Decker2020,Liu2020}. Moreover, such expansions can be constructed with symbolic Hamiltonians. Hence, they readily apply to many-body quantum systems despite their high-dimensional Hilbert space, which makes explicit solutions of the system dynamics practically impossible.

A disadvantage of this generally successful method is its perturbative character, which requires very large frequencies and/or strong driving in order to obtain sizable higher-order interactions. The central goal of this Letter is to overcome this limitation and to arrive at effective systems with three-body interactions that exceed in strength all other interactions, using weak driving only.

This objective is reached by first finding driving patterns that induce exact, desired dynamics in small subsystems of the full system. These building blocks are then concatenated in order to realize the desired dynamics in the full system, be it exactly or approximately. Since each such subsystem needs to contain only $n$ constituents in order to realize effective $n$-body interactions, these subsystems can in practice be as small as three- or four-body systems, and can thus be treated in a (numerically) exact manner.

While the methodology developed in this Letter applies to the full range of driven, composite quantum systems,
we illustrate its properties with the example of an Ising chain with additional driving of single-spin dynamics.
The resulting effective Hamiltonian has a dominant three-body interaction,
with applications in
quantum information \cite{Raussendorf2003,Doherty2009} and condensed-matter physics \cite{Smacchia2011,Son2011,Pachos2004},
and can be readily realized
in state-of-the-art platforms  such as superconducting circuits \cite{Gu2017,Wendin2017} or molecular nanomagnets \cite{Gatteschi2006,Luis2014}, as discussed below.

According to Floquet theory~\cite{Shirley1965,Tannor2007}, the propagator $U(t)$ induced by a periodically time-dependent Hamiltonian $H(t)=H(t+T)$ can be expressed as $U(t)=U_F(t)\exp(-i\Hf t)$, where $U_F(t)$ is periodic with the period $T$ of $H(t)$, and $\Hf$ is a time-independent Hermitean operator. This operator can also be understood as the system Hamiltonian
\begin{equation}
\Hf = U_F^\dagger(t)H(t)U_F(t)-iU^{\dagger}_F(t)\dot{U}_F(t)
\end{equation}
in the periodically oscillating frame defined by $U_F(t)$.
 Since $U_F(t)$ reduces to the identity for integer multiples of the period,
$\Hf$ captures the full system dynamics in this coarse-grained  sense, and the operator $\Hf$ thus warrants the name of \emph{Floquet} or {\em effective Hamiltonian}.

A quantum simulation realized in the spirit of Floquet theory is typically referred to as {\em analog}.
This is in contrast to {\em digital} quantum simulations, which are used in cases in which individual terms $H_j$ of a Hamiltonian $H=
\sum_jH_j$ can be implemented, whereas the simultaneous implementation of several of these terms is not feasible.
In this case, the Trotter-Suzuki formula~\cite{Suzuki1976,Georgescu2014,Lloyd1996}
\begin{equation}
\Bigl(\prod_{j} e^{-i H_j\frac{t}{m}}\Bigr)^m  \simeq e^{-iHt} - \frac{t^2}{2m}\sum_{j<k}[H_j,H_k] \label{eq:trotter}
\end{equation}
asserts that subsequent intervals of dynamics produced by the individual terms of the Hamiltonian approximate the dynamics generated by $H$ with an error due to the non-commutativity of the $H_j$ that scales $\propto t^2/m$.

Since the construction of time-dependent Hamiltonians that result in a desired effective Hamiltonian $\Hf$ for a many-body system is practically only possible with perturbative methods, such as a high-frequency expansion, we will strive for a hybrid of analog and digital quantum simulation.

\begin{figure}[t!]
\centering
\includegraphics[width=1.1\linewidth]{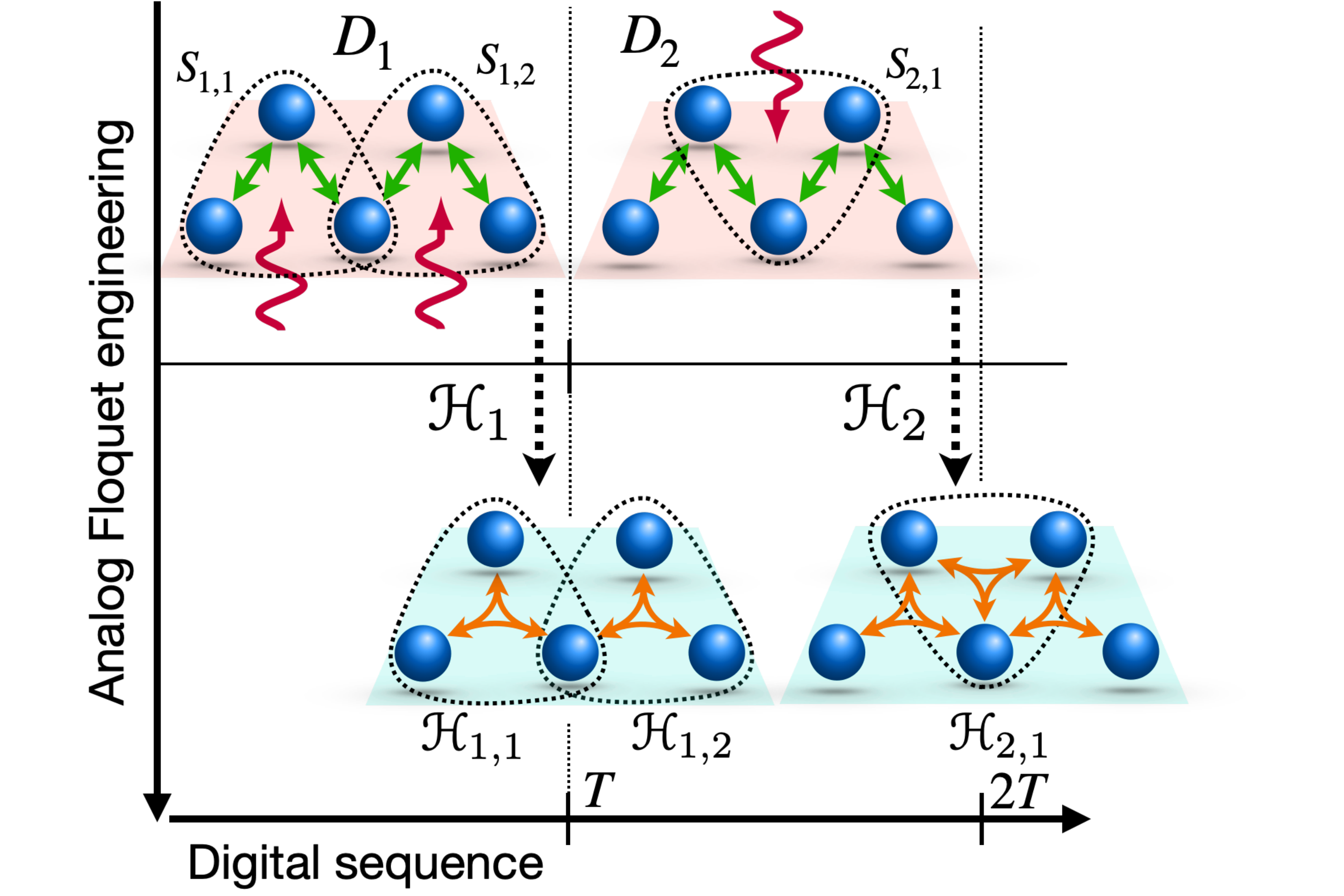}
\caption{A many-body system is decomposed into series of three-body systems in different ways.
For each of the different decompositions $D_j$, the system can be driven such that each three-body subsystem evolves according to a desired effective Hamiltonian.
A concatenation of these different driving protocols in the sense of a digital quantum simulation then results in the dynamics associated with a fully interacting effective Hamiltonian.}
\label{fig1}
\end{figure}

Each of the digital steps of duration $T$ is associated with a decomposition $D_j$ of the full system into several subsystems $S_{j,k}$ as sketched in Fig.~\ref{fig1}.
The decompositions $D_j$ are chosen such that the corresponding propagator factorizes into a product of propagators $U_{j,k}=\exp(-i\Hf_{j,k}T)$ with the effective Hamiltonians $\Hf_{j,k}$ pertaining to the individual subsystems $S_{j,k}$. 
The first goal is to find periodic driving patterns for the subsystems such that each $S_{j,k}$ evolves according to an effective Hamiltonian containing the desired $n$-body interactions (vertical axis in Fig.~\ref{fig1}).
Simultaneous driving of all subsystems in decomposition $D_j$ results then in the effective Hamiltonian $\Hf_j =\sum_k \Hf_{j,k}$ with $n$-body interactions in accordance with the decomposition $D_j$.
After a series of digital steps (horizontal axis in Fig.~\ref{fig1}) where each decomposition is driven $m$ times for a duration of $t/m$, the propagator $U(t)$ for the full system is then given in leading order by the full effective Hamiltonian $\Hf=\sum_{j}\Hf_j$ that realizes the desired many-body dynamics; deviations from this desired dynamics scale as $t^2/m$ following Eq.~\eqref{eq:trotter}.

The central benefit of this approach lies in the fact that the subsystems $S_{j,k}$ are typically sufficiently small so that the driving patterns (or suitable time-dependent Hamiltonians) producing the target Hamiltonian on $S_{j,k}$ can be constructed based on the numerically exact dynamics, without resorting to perturbative schemes.
To this end, it is essential to choose the decompositions
such that driving within a given subsystem  $S_{j,k}$ 
does not impact the dynamics of any other such subsystem $S_{j,\ell}$.
This is automatically ensured if the driven Hamiltonians $H_{j,k}(t)$ of the subsystems $S_{j,k}$ are mutually commuting, {\it i.e.} $[H_{j,k}(t),H_{j,\ell}(t)]=0$. If this is not the case, the decompositions $D_j$ are chosen such that different subsystems can be effectively decoupled with local control fields that make interactions between subsystems far-off-resonant, as illustrated in the SM~\cite{SM}.
In both cases, it is possible to identify suitable control Hamiltonians in terms of the dynamics within the respective subsystem only.

A class of target Hamiltonians for which this strategy is particularly efficient is that of stabilizer Hamiltonians, {\it i.e.}, Hamiltonians defined as a sum of commuting terms. These Hamiltonians are widespread in the fields of quantum computation, quantum error correction, and topological physics \cite{Kitaev2003,Terhal2015,Raussendorf2003,Doherty2009}.
For such a class, the effective Hamiltonian can be decomposed into decompositions $D_j$, such that the Hamiltonians $\Hf_j$ commute.
In this case, there are no additional errors due to the digital steps of the quantum simulation,
and the precision of the Floquet engineering in the single decompositions is maintained on the whole system.

The methodology can be illustrated with a one-dimensional spin chain with Ising interaction
\begin{equation}\label{eq:Ising}
H_{I} = J \sum_{k}\hat{Z}_k \hat{Z}_{k+1}\ ,
\end{equation}
where $\hat{X}_k, \hat{Y}_k$, and $\hat{Z}_k$ are the Pauli matrices for the spin with index $k$.
The goal lies in realizing dynamics associated with the three-body Hamiltonian,
\begin{equation}\label{eq:clust_ham}
H_{\textrm{zxz}} = J_{\textrm{zxz}} \sum_{k} \hat{Z}_{k-1} \hat{X}_{k} \hat{Z}_{k+1}\ ,
\end{equation}
which represents a fundamental model in quantum computation and condensed-matter physics. The ground state of the system with periodic boundary conditions is a cluster state that enables measurement-based quantum computation \cite{Raussendorf2003},
and the open-chain variant of this Hamiltonian hosts a symmetry-protected topological phase \cite{Son2011,Smacchia2011}. 

The central building block for the following construction is a system of $N=3$ spins governed by $H_I+H_c(t)$, with the control Hamiltonian $H_c(t)=f(t) \hat{X}_2$ driving the central spin and a to-be-determined function $f(t)$.

\begin{figure}
 \centering
\includegraphics[width=\linewidth]{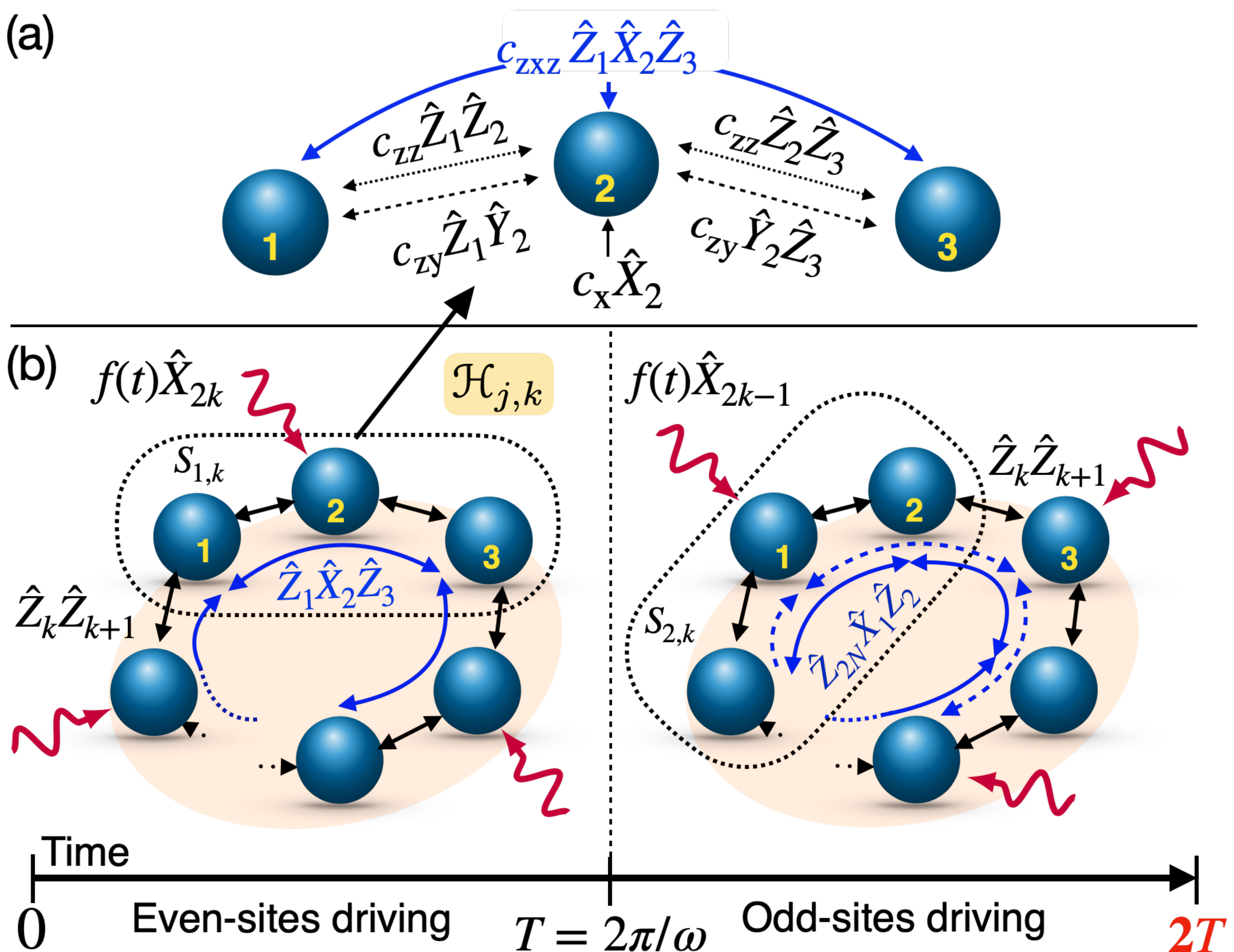}
\caption{(a) Structure of the effective Hamiltonian for a three-spin subsystem, resulting from driving the intermediate site.
An optimized driving field $f(t)$ allows one to isolate the effective three-body term in Eq. \eqref{eq:effHam}, making the others vanish. (b) Two-step procedure for generating the target Hamiltonian $H_{\textrm{zxz}}$ on the full system: even- and odd-labeled spins are driven alternatively, realizing $H_{\textrm{zxz}}$ every second driving period. Dotted panels highlight the fundamental three-spin units featuring the effective Hamiltonian of (a).
}
\label{fig2}
\end{figure}

As discussed in more detail in the SM~\cite{SM}, the range of effective Hamiltonians that can be realized in this fashion is fundamentally restricted to linear combinations of nested commutators of $\hat{X}_2$ and $\hat{Z}_1\hat{Z}_2+\hat{Z}_2\hat{Z}_3$.
The effective Hamiltonian is thus of the form
\begin{multline}\label{eq:effHam}
 c_{\textrm{x}} \hat{X}_2 +  c_{\textrm{zz}}(\hat{Z}_1 \hat{Z}_{2} + \hat{Z}_2 \hat{Z}_{3}) \\
   + c_{\textrm{zy}}(\hat{Z}_1 \hat{Y}_2 + \hat{Y}_2 \hat{Z}_3) 
 +  c_{\textrm{zxz}} \hat{Z}_1\hat{X}_2\hat{Z}_3\ ,
\end{multline}
with the real scalars $c_{\textrm{x}}$, $c_{\textrm{zz}}$, $c_{\textrm{zy}}$ and $c_{\textrm{zxz}}$, as depicted in Fig.~\ref{fig2}a.
Their dependence on the driving pattern $f(t)$ is hopelessly complicated, but it is ensured~\cite{Dalessandro2007,SM} that any effective Hamiltonian of the form in Eq.~\eqref{eq:effHam} can be obtained through the choice of a suitable driving function $f(t)$ and time interval $T$.

A Fourier sum $f(t) = - c_{\textrm{zxz}} + f_1 \cos\omega t + f_2 \cos(2 \omega t)$ with two tuneable parameters $f_i$ is enough to find good solutions via numerical optimisation~\cite{SM}.
Table \ref{tab:table1} exemplifies optimal values for the Fourier coefficients $f_i$ for different values of the driving frequency $\omega$.
While the coefficient $c_{\textrm{zxz}}$ is only slightly smaller in strength than the coupling constant $J$ of the undriven system,
all the other coefficients corresponding to undesired processes are zero within the numerical accuracy.
\begin{table}[t]
\caption{\label{tab:table1}%
Examples of optimal driving amplitudes (truncated to four digits) for different driving frequencies $\omega$ and corresponding coefficients of the effective Hamiltonian \eqref{eq:effHam} for a three-spin subsystem --- double-precision values given in \cite{SM}.}
\begin{ruledtabular}
\begin{tabular}{lccccccr}
\textrm{$\omega/J$} &
\textrm{$f_1/\omega$}&
\textrm{$f_2/\omega$} & \textrm{$c_{\textrm{zxz}}/J$}\\ 
\colrule
1 &  $-2.1285$ & $-$2.6782 & $-0.2$ \\
2 &  $-$0.1566 & $-$0.1984 & $-$0.2 \\
5 & 1.0850 & 1.2455  &$-$0.2 
\end{tabular}
\end{ruledtabular}
\end{table}

With the numerically obtained driving pattern, one can now complete the protocol. There will be two different decompositions in which the system is divided into subsystems comprising three spins each.
In decomposition $D_1$ each subsystem comprises a spin with even index and both neighboring spins, while each subsystem in decomposition $D_2$ comprises a spin with odd index and both neighboring spins [Fig.~\ref{fig2}(b)].
In this way, there are spins that contribute to two different subsystems, but this will not result in any complications because the Hamiltonians of different subsystems commute.

Driving every spin with even index with the optimized driving profile realizes the desired interaction $\hat{Z}_{k-1} \hat{X}_k \hat{Z}_{k+1}$ in every subsystem of decomposition $D_1$, resulting in an effective Hamiltonian of the desired form of Eq.~\eqref{eq:clust_ham}, but with the summation restricted to even values of the index $k$ [see Fig.~\ref{fig2}(b)].
Driving every spin with odd index, on the other hand, realizes this type of Hamiltonian with the summation restricted to odd index values.
A sequence of periods in which, in an alternating fashion, all even spins and all odd spins are driven, thus results in the desired effective Hamiltonian [Eq.~\eqref{eq:clust_ham}] through the Trotter decomposition.
Since all involved terms commute, {\it i.e.}, $[\hat{Z}_{j-1}\hat{X}_j \hat{Z}_{j+1},\hat{Z}_{k-1} \hat{X}_k \hat{Z}_{k+1}]=0$ for all $j$ and $k$, this digital construction is indeed exact.
Alternating driving of even and odd spins thus results in a quantum simulation with an accuracy that is limited only by the numerical Floquet engineering,
both for the case of periodic boundary conditions and the case of open boundary conditions in which the end spins are not driven.
Since errors resulting from non-commutativity do not arise in the presently discussed example,
the SM~\cite{SM} also contains the discussion of the  three-body interaction $\hat{X}_{j-1} \hat{Z}_{j} \hat{X}_{j+1} + \hat{Y}_{j-1} \hat{Z}_{j} \hat{Y}_{j+1}$ realized in a system with an intrinsic $XY$ interaction in order to assess the accuracy that is achievable in the presence of Trotterization and subsystem-decoupling errors.

\begin{figure}[t]
\includegraphics[width=\linewidth]{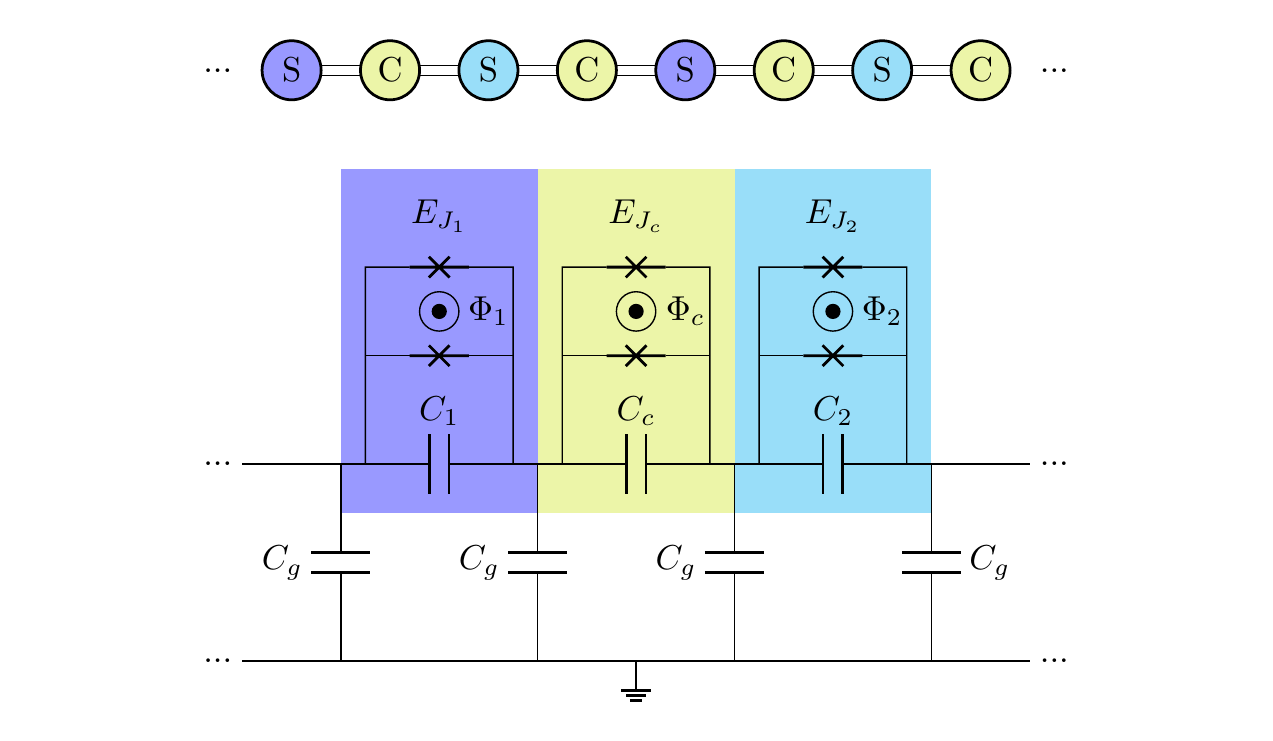}
\caption{Scheme of a superconducting circuit implementing the driven model that simulates $H_{\textrm{zxz}}$. Spins of the model, shown in blue and cyan, are implemented by transmon qubits with Josephson energy $E_{J_k}$ and shunt capacitance $C_k$. Blue and cyan indicate different transition frequencies that can be tuned via external flux $\Phi_k$. The interaction of Eq. \eqref{eq:Ising} is mediated by a nonlinear coupler (C, yellow) with similar circuit as a trasmon qubit. $C_g$ characterizes the capacitive coupling of the chain to the ground.}
\label{fig_x}
\end{figure}

The scheme described here for realizing $H_{\textrm{zxz}}$ can be implemented in a variety of state-of-the-art experimental platforms and two examples are explicitly discussed here. The Ising two-body interaction $H_I$ of Eq. \eqref{eq:Ising} can be obtained in superconducting circuits by exploiting a nonlinear coupling between superconducting qubits \cite{Kounalakis2018,Collodo2019}. A cross-Kerr interaction between transmons can be engendered such that, by parametrically switching off flip-flop terms, it results in a pure $ZZ$ interaction in the qubit subspace \cite{Kounalakis2018} (of a few tens of megahertz in strength). This architecture is represented in Fig.~\ref{fig_x}. Oscillating control fields are achieved by applying a time-dependent voltage $v(t) \propto f(t)\sin(\omega_k t)$ to the drive line of each qubit, where $\omega_k$ is the qubit's transition frequency. Independent qubit control hence permits one to address even and odd sublattices separately. The driven Hamiltonian realizing $H_{\textrm{zxz}}$ effectively is finally obtained in the interaction picture with respect to the bare qubit energies. An initial two-body coupling $J/2\pi= 10$ MHz translates, for a driving frequency $\omega/2\pi=50$ MHz, into a three-body coupling of strength $|J_{\textrm{zxz}}|/2\pi \sim 2$ MHz, with control amplitudes smaller in strength than $63$ MHz. Relaxation and dephasing times around $\tau_c \simeq 20$ $\mu$s for the single qubits \cite{Kounalakis2018} are then around five hundred times longer than the stroboscopic period $2T$, and the strong coupling condition $\tau_c |J_{\textrm{zxz}}|/2\pi  \gg 1$ is indeed given for the effective three-body interaction.

A second promising platform is that of molecular nanomagnets. A suitable molecular architecture is given by heterometallic complexes featuring an alternating arrangement of two types of ions $A$ and $B$ with different $g$ factors. At low temperatures, their magnetic ground state doublets behave as a chain of weakly interacting effective spin-$1/2$ systems, mainly coupled by dipolar magnetic interaction \cite{Aromi2014,Chilton2018,Macaluso2020}. 
A sufficiently strong static magnetic field along $z$ effectively suppresses flip-flop processes between neighboring spins and potential crosstalk errors, thanks to the difference in $g$ factors, leaving an effective Ising interaction only. 
Periodic driving of even and odd subchains is achieved by means of electromagnetic fields oscillating in the $x$-$y$ plane, with tones both resonant with the bare transitions of spins of type $A$ and $B$, respectively, and detuned by $\pm \omega$ and $\pm 2 \omega$ \cite{SM}. The driven Hamiltonian simulating $H_{\textrm{zxz}}$ is then attained in the interaction picture. Assuming a pairwise coupling $J/2\pi= 75$ MHz and $\omega=5J$, the corresponding parameters in Table \ref{tab:table1} translate, for instance, into amplitudes of the oscillating magnetic fields between 2 and 52 G for realistic $g$ factors \cite{SM}, resulting in an effective three-body coupling $|J_{\textrm{zxz}}|/2\pi=14$ MHz. 
Given typical coherence times of several microseconds for these complexes \cite{Wedge2012,Shiddiq2016,Zadrozny2015}, strong coupling is well attained.

The effective Hamiltonian simulating $H_{\textrm{zxz}}$ can be used, for instance, for preparing the cluster state $\ket{c}$. This state can be defined in a compact way as the action of controlled-$Z$ (CZ) operations on any two neighboring spins, starting from an initial state which is a tensor product of the +1 eigenvector of $\hat{X}_k$, here denoted with $\ket{+_k}$, for all spins. In formula,
\begin{equation}
\ket{c} = \mathrm{CZ}_{1,2}\cdot \ldots \cdot \mathrm{CZ}_{N-1,N} \mathrm{CZ}_{N,1} \ket{+_1\ldots +_N }.
\end{equation}
The state preparation can be done by interpolating between the initial static Hamiltonian and the effective Hamiltonian, such that the ground state of the first is adiabatically connected to $\ket{c}$ \cite{Eckardt2017}. In practice, this implies ramping up the periodic drive while still addressing even-odd sites alternatively, in a time window $t_f$. 
For instance, starting at time $t=0$ with the system in the ground state of $H(t) = \sum_{k} [ h(t) \hat{X}_k + \hat{Z}_k \hat{Z}_{k+1}]$ for $h(t) = [1-r(t/t_f)]+r(t/t_f) f(t)$ for six spins, the state preparation is achieved with fidelity above 0.99 for total ramping time $t_f>250T$, using optimal driving parameters $\omega=10J$, $c_{\mathrm{zxz}}/\omega=-0.009$, $f_1/\omega=1.200$, $f_2/\omega=1.224$ and 
a polynomial ramp $r(t) =  35 t^4 - 84 t^5 + 70 t^6 - 20 t^7$ with three vanishing time derivatives at the boundary \cite{Petiziol2019}. 

The technique presented in this Letter can be applied to a plethora of many-body models, and the effective Hamiltonians thus achieved can be exploited both for studying fundamental physics and for the preparation of quantum states with interesting properties. 
For instance, the exemplary construction of the Hamiltonian $H_{\textrm{zxz}}$ can be adapted to study the cluster phase transition \cite{Pachos2004}. Indeed, this can be achieved by varying the static component $c_{\textrm{zxz}}$ in the driving profile $f(t)$.
Similarly, the present strategy can be exploited to simulate a cluster-Ising phase transitions \cite{Son2011,Smacchia2011}, or to facilitate the experimental accessibility of recently-proposed models where Floquet driving, many-body localization and disorder coexist \cite{Potirniche2017,Decker2020}. The efficiency of the method in the simulation of stabilizer Hamiltonians may further pave the way to the investigation and realization of Floquet-engineered self-correcting quantum memories \cite{Terhal2015}, for which the encoded logical space corresponds to the low-energy sector of an effective Hamiltonian engineered through the driving. Concerning the simulation of $n$-body interactions for $n>3$, the analog-digital strategy can be applied straightforwardly. While the promotion of selected $(n>3)$-body processes from two-body interactions will be accompanied by the creation of an increasing number of undesired terms, many processes can be fully inhibited by specific choices of the driving phases, hence revealing a central role of time symmetries in the drive at higher orders.

\acknowledgments

This collaboration was enabled with financial support from the Imperial College European Partners Programme and the Leonardo da Vinci mobility program promoted by CRUI/MIUR. Financial support from EPSRC through the grant “Optimal Control for Robust Ion Trap Quantum Logic” EP/P024890/1 is gratefully acknowledged. The projects Theory-Blind Quantum Control TheBlinQC, cofunded by EPSRC under the Grant No. EP/R044082/1, and “Scaling Up quantum computation with MOlecular spins” (SUMO), cofunded by the Italian MUR, have received funding from the QuantERA ERANET Cofund in Quantum Technologies implemented within the European Union Horizon 2020 Programme. 


%
%
\clearpage

\onecolumngrid 
\begin{center}
{\bf \large Supplemental Material to} \\
\vspace{0.2cm}
{\bf \large Quantum Simulation of Three-Body Interactions in Weakly Driven Quantum Systems} \\
\vspace{0.4cm}

{ Francesco Petiziol,$^{1,2}$ Mahdi Sameti,$^3$ Stefano Carretta,$^{1,2}$ Sandro Wimberger,$^{1,4}$ and Florian Mintert$^3$ }\\

\vspace{0.3cm}

{\itshape
$^1$Universit\`a di Parma, Dipartimento di Scienze Matematiche, Fisiche e Informatiche, I-43124 Parma, Italy\\
$^2$UdR Parma, INSTM, I-43124 Parma, Italy \\
$^3$Blackett Laboratory, Imperial College London, London SW7 2AZ, United Kingdom \\
$^4$ INFN, Sezione di Milano Bicocca, Gruppo Collegato di Parma, Parma, Italy
}
\end{center}

\vspace{0.5cm}

The Supplemental Material is structured as follows. In Sec. \ref{sec:effham}, details are provided on the derivation of Eq. (5), on the choice of the {\it ansatz} for the control Hamiltonian and control function, on the numerical determination of optimal driving parameters and on the robustness of the effective Hamiltonian with respect to imperfections in these parameters. In Sec. \ref{sec:exp}, the derivation of the driven Hamiltonian simulating $H_{\textrm{zxz}}$ is given for the experimental platform of molecular nanomagnets. In this context, the impact of crosstalk among spins of different species on the desired effective Hamiltonian is quantitatively analysed. In Sec. \ref{sec:xymodel}, an example is discussed in which a non-stabilizer three-body Hamiltonian is simulated starting from non-commuting two-body interactions. This example aims at clarifying the procedure of subsystem decoupling via control fields, the related errors, and the impact of digitization errors on the effective Hamiltonian. 

\section{Effective Hamiltonian for a three-spin unit} \label{sec:effham}

\subsection{Algebraic structure of the effective Hamiltonian}

According to the Lie Algebra Rank Condition of quantum controllability theory~\cite{Dalessandro2007}, the propagator $U(t)$ induced by a generic driven Hamiltonian of the form $H=H_0+\sum_j f_j(t)H_j$ with time-dependent functions $f_j(t)$ is of the form
\begin{equation}
U(t)=\exp\left(-i\sum_j g_j(t)K_j\right)\ ,
\end{equation}
where the set of operators $K_j$ is comprised of the $H_j$ and their nested commutators, and the $g_j(t)$ are scalar, time-dependent functions.
Moreover, one can always find functions $f_j(t)$ such that the system propagator approximates a desired unitary of the form $U_T=\exp\left(-i\sum_j c_jK_j\right)$ arbitrarily well at some point in time.

In the present case the Hamiltonian reads $H(t)=J H_0+f(t)H_1$ with
\begin{subequations}
\begin{eqnarray}
H_0&=&\hat{Z}_1\hat{Z}_2+\hat{Z}_2\hat{Z}_3\ ,\\
H_1&=&\hat{X}_2\ .
\end{eqnarray}
\end{subequations}
The commutator
\be
[H_1,H_0]=\hat{Z}_1[\hat{X}_2,\hat{Z}_2]+[\hat{X}_2,\hat{Z}_2]\hat{Z}_3=-2i(\hat{Z}_1\hat{Y}_2+\hat{Y}_2\hat{Z}_3)=-2iH_2
\ee
defines a third operator $H_2$ that is linearly independent from $H_0$ and $H_1$.
The commutator of $H_1$ and $H_2$ does not add an independent operator, because $[H_1,H_2]=2iH_0$, but the commutator
\be
[H_0,H_2]=
\hat{Z}_1^2[\hat{Z}_2,\hat{Y}_2]+2\hat{Z}_1[\hat{Z}_2,\hat{Y}_2]\hat{Z}_3+[\hat{Z}_2,\hat{Y}_2]\hat{Z}_3^2=
-4i \hat{X}_2-4i\hat{Z}_1\hat{X}_2\hat{Z}_3=-4iH_1-4iH_3
\ee
results in an independent operator $H_3=\hat{Z}_1\hat{X}_2\hat{Z}_3$.
All commutators $[H_j,H_k]$ with $j,k=0,\hdots,3$ can be expressed as linear combinations of the $H_j$,
so that, indeed the propagator induced by the Hamiltonian 
$H(t)= J H_0 + f(t) H_1$ can be written as
\be
U(t)=\exp\left(i\sum_{j=0}^3\eta_j(t)H_j\right)\ ,
\ee
with real, time-dependent functions $\eta_j(t)$. Hence, the effective Hamiltonian can be generally expressed as per Eq. (5) of the main text.

\subsection{Choice of the control function}\label{sec:contrfun}

Complete, analytic solutions for the dynamics induced by the Hamiltonian $H=J(\hat{Z}_1\hat{Z}_2+\hat{Z}_2\hat{Z}_3)+f(t)\hat{X}_2$ with a time-dependent function are practically impossible to find,
but the problem can be simplified since the Hamiltonian commutes with the operators $\hat{Z}_1$ and $\hat{Z}_3$.
The $Z$-components of the first spin and of the third spin are thus conserved,
and the problem can be formulated in terms of the dynamics of the second spin only, with a Hamiltonian that is conditioned on the states of the first and third spin.

Denoting the Hamiltonians of the second spin, conditioned on both other spins to be in their $Z=1$ eigenstates, in their $Z=-1$ eigenstates, or in different $\hat{Z}$ eigenstates by $H_+$, $H_-$ and $H_0$,
the dynamics of the three-spin system is completely solved in terms of the dynamics induced by the Hamiltonians
\begin{subequations}
\begin{eqnarray}
H_+&=&f(t)\hat{X}_2+2 J \hat{Z}_2\ ,\\
H_-&=&f(t)\hat{X}_2-2J \hat{Z}_2\ ,\\
H_0&=&f(t)\hat{X}_2\ .
\end{eqnarray}
\end{subequations}
Following the same reasoning for $\hat{Z}_1\hat{X}_2\hat{Z}_3$, the goal of the present control problem is the identification of a periodic function $f(t)$ such that $H_0$ induces a propagator satisfying
\begin{equation}
U_0(T)=\exp(ic_{\textrm{zxz}} \hat{X}_2 T)\ ,
\label{eq:cond}
\end{equation}
at the end of the period $T$, with the desired strength $c_{\textrm{zxz}}$ of the three-body interaction $\hat{Z}_1\hat{X}_2\hat{Z}_3$,
while the propagators induced by $H_+$ and $H_-$ satisfy $U_+(T)=U_-(T)=\exp(-ic_{\textrm{zxz}} \hat{X}_2 T)$.

\subsubsection{Fourier sum}

Eq.~\eqref{eq:cond} implies that $\int_{0}^{T}dt\ f(t)= -c_{\textrm{zxz}}T$, which shows that a parametrization of $f(t)$ as a Fourier sum requires a static component,
and its value is determined by the targeted interaction constant.
The prefactor $c_{\mathrm{zy}}$ of the term $\hat{Z}_1\hat{Y}_2+\hat{Y}_2\hat{Z}_3$ in the effective Hamiltonian (Eq.(5)) necessarily vanishes for driving functions $f(T/2+t)=f(T/2-t)$ that are symmetric around the center of the interval $[0,T]$ \cite{Brinkmann2016}.

A suitable parametrization for $f(t)$ is thus given in terms of the Fourier sum
\begin{equation}
\label{eq:contr_f}
f(t)=-c_{\textrm{zxz}}
+\sum_{j>0}f_j\cos\left(2\pi j\frac{t}{T}\right)\ .
\end{equation}
The number of tuneable amplitudes $f_j$ required to make sure that also $c_{\mathrm{x}}$ and $c_{\mathrm{zz}}$ vanish in the effective Hamiltonian (Eq.(5)) cannot be determined from first principles,
but we found the two terms are sufficient.

Because of the non-commutativity of $\hat{X}_2$ and $\hat{Z}_2$, the dynamics induced by $H_+$ and by $H_-$ can not be solved for general driving functions $f(t)$,
but suitable amplitudes $f_j$ can be obtained based on numerical solution of the Schr\"odinger equation, as discussed below.

\subsection{Numerical determination of the driving amplitudes} \label{sec:optim}

The driving function $f(t)$ can be optimized based on a numerically constructed propagator $U(T)$,
which can be obtained, for example, through integration of the Schr\"odinger equation or diagonalisation of the Floquet operator~\cite{Bartels2013}.
The quantity to minimize would be the gate-fidelity $\Vert U(T)-\exp(-iH_{\mathrm{tg}}T) \Vert$, where $\Vert \cdot \Vert$ denotes here the Hilbert-Schmidt norm and $H_{\mathrm{tg}}$ denotes the targeted, effective Hamiltonian. In the present case, the latter is $c_{\textrm{zxz}}\hat{Z}_1\hat{X}_2\hat{Z}_3$.
The determination of the propagator $U(T)$ for a subsystem of $n$ spins requires in general to solve the dynamics of the system of $n$ spins. However, the decomposition into three independent single-qubit problems and the identification of conditions on the driving function $f(t)$ discussed above, results in a simplified numerical construction for the specific model considered here (see Sec. \ref{sec:xymodel} for a different case).

Equation~\eqref{eq:cond} is naturally satisfied for any driving function in Eq.~\eqref{eq:contr_f},
and since $\hat X_2H_+\hat X_2=H_-$, the propagators $U_+(t)$ and $U_-(t)$ satisfy the relation $\hat X_2U_+(t)\hat X_2=U_-(t)$.
That is, once the condition $U_+(T)=\exp(-ic_{\textrm{zxz}} \hat{X}_2 T)$ is satisfied, the condition $U_-(T)=\exp(-ic_{\textrm{zxz}} \hat{X}_2 T)$ is satisfied as well.
The remaining optimization problem to be solved is thus the problem of a single, poly-chromatically driven qubit.

The propagator induced by $H_+$ with a driving function $f(t)$ that is symmetric around $T/2$ can be written in the form
\be
U_+(T)=\exp\left(-i\left((d_x+c_{\textrm{zxz}})\hat X_2+d_z\hat Z_2\right)T\right)\ ,
\label{eq:Uplus}
\ee
for some scalars $d_x$ and $d_z$, but without $\hat{Y}_2$-term.
The quest for an explicit function $f(t)$ that results in a propagator with vanishing $d_x$ and $d_z$ can be performed numerically.
Due to numerical precision of such an optimization,
the final values $d_x$ and $d_z$ will not be strictly zero, resulting in finite values of the parameters $c_{\mathrm{x}}$ and $c_{\mathrm{zz}}$ in the effective Hamiltonian for the three-spin system.

In order to relate the values of the scalars $d_x$  and $d_z$ to the precision of the resulting effective Hamiltonian of the three-spin system ({\it i.e.}, to the coefficients $c_{\mathrm{x}}$, $c_{\mathrm{zz}}$ and $c_{\mathrm{zxz}}$ of Eq.~(5)), it is instructive to apply the general form
\be
U(T)=\exp\left[-i\left(c_{\textrm{x}}\hat{X}_2+
c_{\textrm{zz}}(\hat{Z}_1\hat{Z}_2+\hat{Z}_{2}\hat{Z}_3)+
c_{\textrm{zxz}}\hat{Z}_1\hat{X}_2\hat{Z}_3\right)T\right]
\ee
of the propagator [stated in Eq.(5) in the main paper with extra $(\hat{Z}_1\hat{Y}_2+\hat{Y}_{2}\hat{Z}_3)$-term]
to the state $\ket{\uparrow\Psi\uparrow}$ of a three qubit system.
Since for this initial state the operators $\hat{Z}_1$ and $\hat{Z}_3$ can be replaced by their eigenvalue $1$,
this yields
\bqa
U(T)\ket{\uparrow\Psi\uparrow}
&=&\exp\left[-i\left(
(c_{\textrm{x}}+c_{\textrm{zxz}})\hat{X}_2+
2c_{\textrm{zz}}\hat{Z}_2
\right)T\right]\ket{\uparrow\Psi\uparrow}.
\eqa
Identification with Eq.~\eqref{eq:Uplus} then results in the relations
\bqa
c_{\textrm{x}}&=&d_x,\\
2c_{\textrm{zz}}&=&d_z,
\eqa
between the numerically obtained paramters $d_x$, $d_z$ and the resulting magnitude of undesired terms in the effective Hamiltonian of the three-spin system.

Examples of the results of the optimisation for different choices of $\omega$ are shown in Table \ref{tab:1}. 
The finite values of $c_{\textrm{zy}}$ (which is strictly vanishing due to the symmetry of $f(t)$) indicates the finite precision of the numerical integrator,
and the residual finite values for $c_{\textrm{x}}$, $c_{\textrm{zz}}$ are due to the finite precision of the optimization routine.
Apart from limited numerical accuracy, however, those values can be taken to be exactly zero.

\begin{table}[b]
\caption{\label{tab:1}%
Double precision values for the diving parameters given in Table 1 of the main text.}
\begin{ruledtabular}
\begin{tabular}{lccccccr}
\textrm{$\omega/J$} &
\textrm{$f_0/\omega$}&
\textrm{$f_1/\omega$}&
\textrm{$f_2/\omega$}&
\textrm{$c_{\textrm{x}}/J$} & \textrm{$c_{\textrm{zz}}/J$} & \textrm{$c_{\textrm{zy}}/J$} & \textrm{$c_{\textrm{zxz}}/J$}\\ 
\colrule
1 & 0.2 & $-$2.1285310408898304  & $-$2.678165158151617 &   $10^{-11}$ & $10^{-10}$ & $10^{-12}$ & $-$0.2 \\
2 & 0.1 & $-$0.15660928207162234&  $-$0.1984060069686025 & $10^{-14}$ & $10^{-10}$ & $10^{-13}$ & $-$0.2 \\
5 &  0.04  & 1.0849517026900328 & 1.2455409822710848 & $10^{-10}$ & $10^{-11}$ & $10^{-12}$ &$-$0.2
\end{tabular}
\end{ruledtabular}
\end{table}

\begin{figure}[t]
\includegraphics[width=\linewidth]{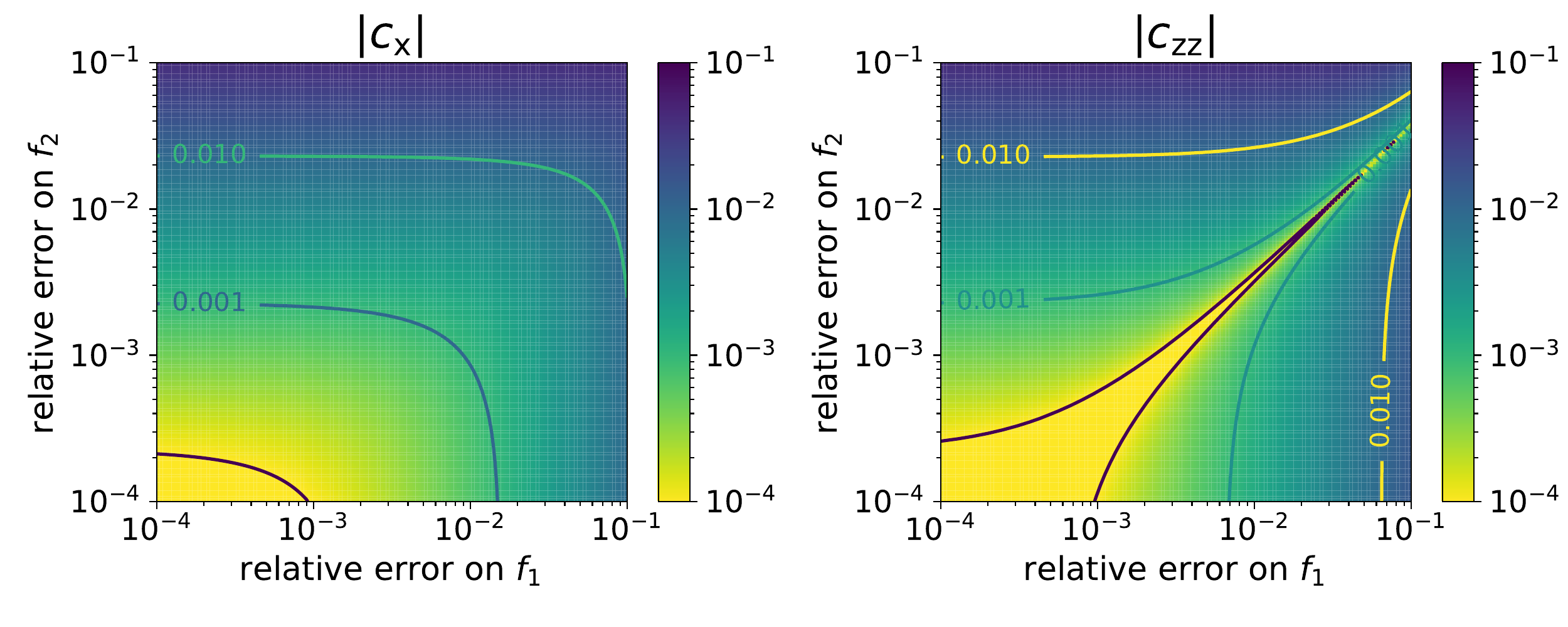}
\caption{Log-color map of the absolute value of the coefficients $c_{\mathrm{x}}$ and  $c_{\mathrm{zz}}$ as a function of the relative error (in log scale) on the driving amplitudes $f_1$ and $f_2$, for $f_0 = -c_{\textrm{zxz}} = 0.04\omega$, $\omega=5J$ and corresponding unperturbed driving parameters given in Table \ref{tab:1}.}
\label{fig:maperrs}
\end{figure}

In order to assess the accuracy in pulse shaping that is required in an experimental implementation,
Fig.~\ref{fig:maperrs} depicts the values for $c_{\mathrm{x}}$ and  $c_{\mathrm{zz}}$ that are obtained with driving that deviates from an optimized driving pattern. The value of the ideally-zero coefficients $c_{\mathrm{x}}$ and  $c_{\mathrm{zz}}$ remains below $10^{-3}$ for relative errors of $10^{-2}$ in $f_1$ and $10^{-3}$ in $f_2$, thus confirming the robustness of the quantum simulation to fluctuations in the driving field.

\subsection{Analogy to spin echo}

While the above parametrization in terms of a Fourier sum results in driving profiles that are practical for an implementation,
one can find another solution that helps to gain a more intuitive understanding of the control problem by drawing an analogy to the concept of the spin echo.
Two intervals of duration $\frac{3\pi}{4J}$ and $\frac{\pi}{4J}$ of un-driven ({\it i.e.} $f(t)=0$) dynamics result in the propagators $\exp( -i\frac{3\pi}{2}\hat{Z}_2)= i \hat{Z}_2$ and $\exp(- i\frac{\pi}{2}\hat{Z}_2)= -i \hat{Z}_2$ in the case of $H_+$, and $\exp( i\frac{3\pi}{2}\hat{Z}_2)= -i \hat{Z}_2$ and $\exp( i\frac{\pi}{2}\hat{Z}_2)= i \hat{Z}_2$ in the case of $H_-$, whereas they result in no dynamics in the case of $H_0$.
A hard and short pulsed drive, results in free $\hat{X}_2$ rotation in all three cases.

With a driving function
\begin{equation}
f(t)=\lambda\ \delta\left(t-\frac{3\pi}{4J}\right)\ ,
\end{equation}
{\it i.e.} a hard pulse between the two intervals of un-driven dynamics,
the three propagators at stake thus satisfy
\begin{subequations}
\begin{eqnarray}
U_+\left(\frac{\pi}{J}\right)&=&\hat{Z}_2\exp(-i\lambda\hat{X}_2)\hat{Z}_2=\exp(i\lambda\hat{X}_2)\ ,\\
U_0\left(\frac{\pi}{J}\right)&=&\hat{\mathbb{1}}_2\exp(-i\lambda\hat{X}_2)\hat{\mathbb{1}}_2=\exp(-i\lambda\hat{X}_2)\ ,\\
U_-\left(\frac{\pi}{J}\right)&=&\hat{Z}_2\exp(-i\lambda\hat{X}_2)\hat{Z}_2=\exp(i\lambda\hat{X}_2)\ ;
\end{eqnarray}
\end{subequations}
that is, the goal is achieved in a time interval $T = \frac{\pi}{J}$.

\section{Driven Hamiltonian for experimental realizations} \label{sec:exp}

In this Section, supplemental details are provided on the mapping of the model for implementing $H_{\textrm{zxz}}$ on an architecture of molecular nanomagnets. A similar treatment applies to the case of superconducting circuits.

\subsection{Hamiltonian and control fields}\label{sec:exper}

 Considering a heterometallic complex with structure [$A$ $B$ $A$ $\dots$] as described in the main text, a chain of three effective electronic spin 1/2 particles representing a three-spin unit of the model is described by the spin Hamiltonian
\begin{equation}\label{eq:Ham}
H =  g_A^z \mu_B B_z (\hat{S}_1^z + \hat{S}_3^z) + g_B^z \mu_B B_z \hat{S}_2^z + J_{\parallel} \sum_{k=1}^2 \hat{S}_k^z \hat{S}_{k+1}^z + J_{\perp} \sum_{k=1}^2 (\hat{S}_k^+\hat{S}_{k+1}^- + \hat{S}_k^- \hat{S}_{k+1}^+) ,
\end{equation}
 where $\{\hat{S}_k^{x},\hat{S}_k^{y},\hat{S}_k^{z}\} = \{\hat{X}_k/2, \hat{Y}_k/2, \hat{Z}_k/2  \}$, $\hat{S}_k^\pm = \hat{S}_k^x + i \hat{S}_k^y $ are spin-1/2 operators for the $k$-th spin and the values of $g_A^z$, $g_B^z$ and $B_z$ are such that $|(g_A^z-g_B^z) \mu_B B_z |\gg |J_\perp|$. This condition implies that the difference in transition frequency between spins $A$ and $B$ is much larger than the transverse component of the interaction. Thus, it ensures that flip-flop terms between neighbouring spins are strongly unfavored energetically and hence suppressed, such that a pure $\hat{S}_k^z \hat{S}_{k+1}^z$ interaction is effectively  achieved. Moreover, a large value of $|(g_A^z-g_B^z) B_z|$ also reduces crosstalk effects between spins of type $A$ and $B$, as discussed more in detail in the following.
The periodic driving is implemented through magnetic fields oscillating in the $x$-$y$ plane. The driving Hamiltonian reads
\begin{equation}\label{eq:driving}
H_d(t) = g_A^x \mu_B B_x(t) [\hat{S}_1^x + \hat{S}_3^x] + g_A^y \mu_B B_y(t) [ \hat{S}_1^y +  \hat{S}_3^y] + g_B^x \mu_B B_x(t) \hat{S}_2^x + g_B^y \mu_B B_y(t) \hat{S}_2^y,
\end{equation}
with $B_x(t)$ and $B_y(t)$ of the form
\begin{subequations}\label{eq:magfields}
\begin{align}
B_x(t) & = B_0^x \cos(\Omega_2 t)  + \sum_{\alpha=1}^{2} B_\alpha^x\{\cos[(\Omega_2 + \alpha \omega) t] + \cos[(\Omega_2-\alpha\omega)t]\}, \\
B_y(t) & = B_0^y \sin(\Omega_2 t) + \sum_{\alpha=1}^{2} B_\alpha^y\{\sin[(\Omega_2 + \alpha \omega) t] + \sin[(\Omega_2-\alpha\omega)t]\},
\end{align}
\end{subequations}
where $\Omega_2 = g_B^z \mu_B B_z/\hbar$ is the bare transition frequency of the central spin. Therefore, the magnetic control fields are composed of oscillating fields with one oscillating component which is resonant with the bare transition of the central spin, and other oscillating components symmetrically detuned by $\pm \omega$ and $\pm 2 \omega$. In the interaction picture with respect to the non-interacting part of Eq. \eqref{eq:Ham}, defining $\hbar J\equiv J_{\parallel}/4$ and choosing $B_0^{x,y}= 2 \hbar f_0/ g_B^{x,y}\mu_B$, $B_{1}^{x,y} = \hbar f_{1}/g_B^{x,y} \mu_B $, $B_{2}^{x,y} = \hbar f_{2}/g_B^{x,y} \mu_B $, the Hamiltonian
\begin{equation}
 \hbar f(t) \hat{X}_2 +  \hbar J[\hat{Z}_1 \hat{Z}_2 + \hat{Z}_2 \hat{Z}_3],
\end{equation}
of the model for a three-spin unit is obtained, where terms rapidly oscillating at frequencies equal to or larger than  $(g_A^z - g_B^z) \mu_B B_z/\hbar$ have been neglected.
As an example, let us translate the optimal parameters corresponding to $\omega=5J$ in Table \ref{tab:1} to the molecular system.  
For $\bm{g}_A = \{14.2,3.2,0.5 \}$ and $\bm{g}_B = \{9.3,6.4,4.0 \}$ \cite{Chilton2018} and assuming a value $J_{\parallel}/(2\pi\hbar) = 0.3$ GHz, the detuning $\omega$ is then equal to $\omega/2\pi \simeq 375$ MHz, while the magnetic field amplitudes are 
$B_0^x = 2.3$ G, $B_1^x = 32.3$ G , $B_2^x = 35.9$ G, $B_0^y = 3.3$ G, $B_1^y = 45.4$ G , $B_2^y = 52.1$ G.
 The resulting three-body interaction $-\hbar J_{\textrm{zxz}} \hat{Z}_1\hat{X}_2 \hat{Z}_3$ has strength $J_{\textrm{zxz}}/2\pi \simeq 14$ MHz.

\subsection{Impact of spin cross-talk}
\label{sec:crosstalk}

\begin{figure}
\includegraphics[width=0.5\textwidth]{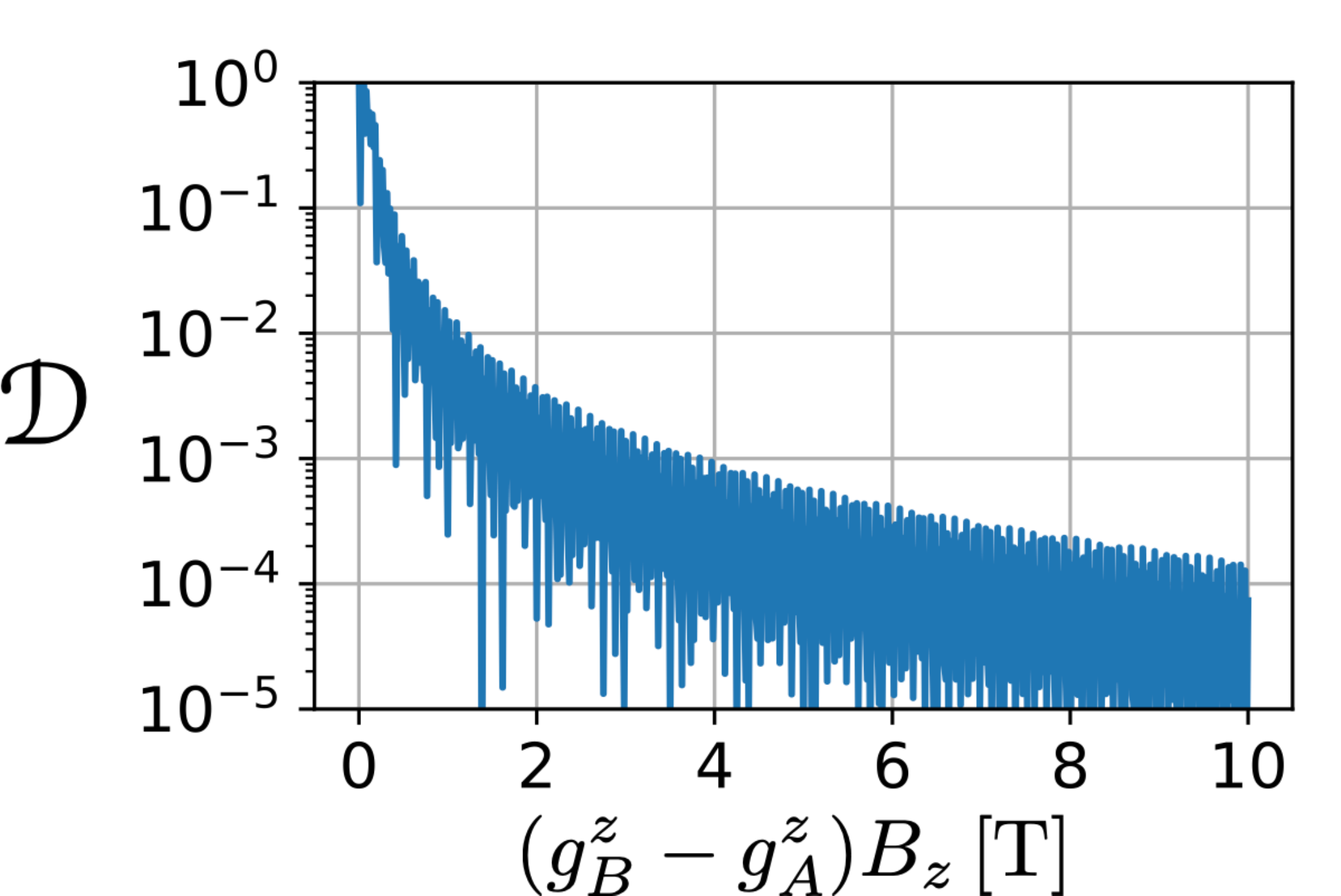}
\caption{Relative distance $\mathcal{D}$ between the effective Hamiltonians in the presence and in the absence of spin crosstalk induced by the driving, as a function of the product $(g_B^z - g_A^z) B_z$, with $g_B^z>g_A^z$. The effective Hamiltonians correspond to parameters $\bm{g}_A = \{14.2,3.2,0.5 \}$, $\bm{g}_B = \{9.3,6.4,g_B^z \}$ with $g_B^z$ being varied, $J_{\parallel}/2\pi \hbar= 0.3$ GHz, $\hbar \omega = 5 J_\parallel/4$ and driving amplitudes given at the end of Sec. \ref{sec:exper}.}
\label{fig:crosstalk}
\end{figure}

The oscillating magnetic fields used for control address all spins at the same time as per Eq. \eqref{eq:driving} and can thus lead to crosstalk between spins of type $A$ and $B$. Selective control of spins of type $A$ and $B$ is achieved by setting the magnetic fields in resonance (and slightly detuned by $\omega$ and $2\omega$, see Eq. \eqref{eq:magfields}) with spins $A$ or $B$, and by ensuring a large difference in transition frequencies, which in turn requires a large value of the product $|(g_A^z-g_B^z) B_z|$.
Therefore, crosstalk can be reduced by considering species with a larger difference in $g^z$ factors , and/or by using a stronger static field $B_z$. For the parameters considered here, it also holds that $2\omega \ll  \mu_B |(g_A^z-g_B^z) B_z|/\hbar$, thus guaranteeing that the detuning from $\Omega_2$ is small with respect to the gap in transition frequency of spins $A$ and $B$, {\it i.e.}, also detuned components of the drive are off-resonant with respect to spins $A$. To verify quantitatively that crosstalk does not impact significantly the desired effective three-body Hamiltonian, Fig. \ref{fig:crosstalk} depicts the relative distance,
\begin{equation}\label{eq:norms}
\mathcal{D} = \left\lvert 1 - \frac{\text{tr}[ \he\, \he^{(0)}]}{\text{tr}[(\he^{(0)})^2]} \right\lvert,
\end{equation}
between the ideal effective Hamiltonian $\mathcal{H}^{(0)}$ and the actual effective Hamiltonian $\mathcal{H}$ as function of the product $(g_B^z-g_A^z) B_z$. The ideal effective Hamiltonian  $\mathcal{H}^{(0)}$ is obtained by completely removing the off-resonant coupling of the control fields to the spins, \emph{i.e.}, the first and second terms in Eq. \eqref{eq:driving}, whereas the actual effective Hamiltonian $\mathcal{H}$ is obtained from the full Hamiltonian in Eq. \eqref{eq:driving}.
One can appreciate that, for the product $|(g_A^z-g_B^z) B_z|$ larger than $\sim 1$~T, the effective Hamiltonian differs by less than 1\% from the ideal one, thus confirming that spin crosstalk is not a serious limitation.

\section{Non-commuting interactions}\label{sec:xymodel}

\begin{figure}[t]
\includegraphics[width=0.7\textwidth]{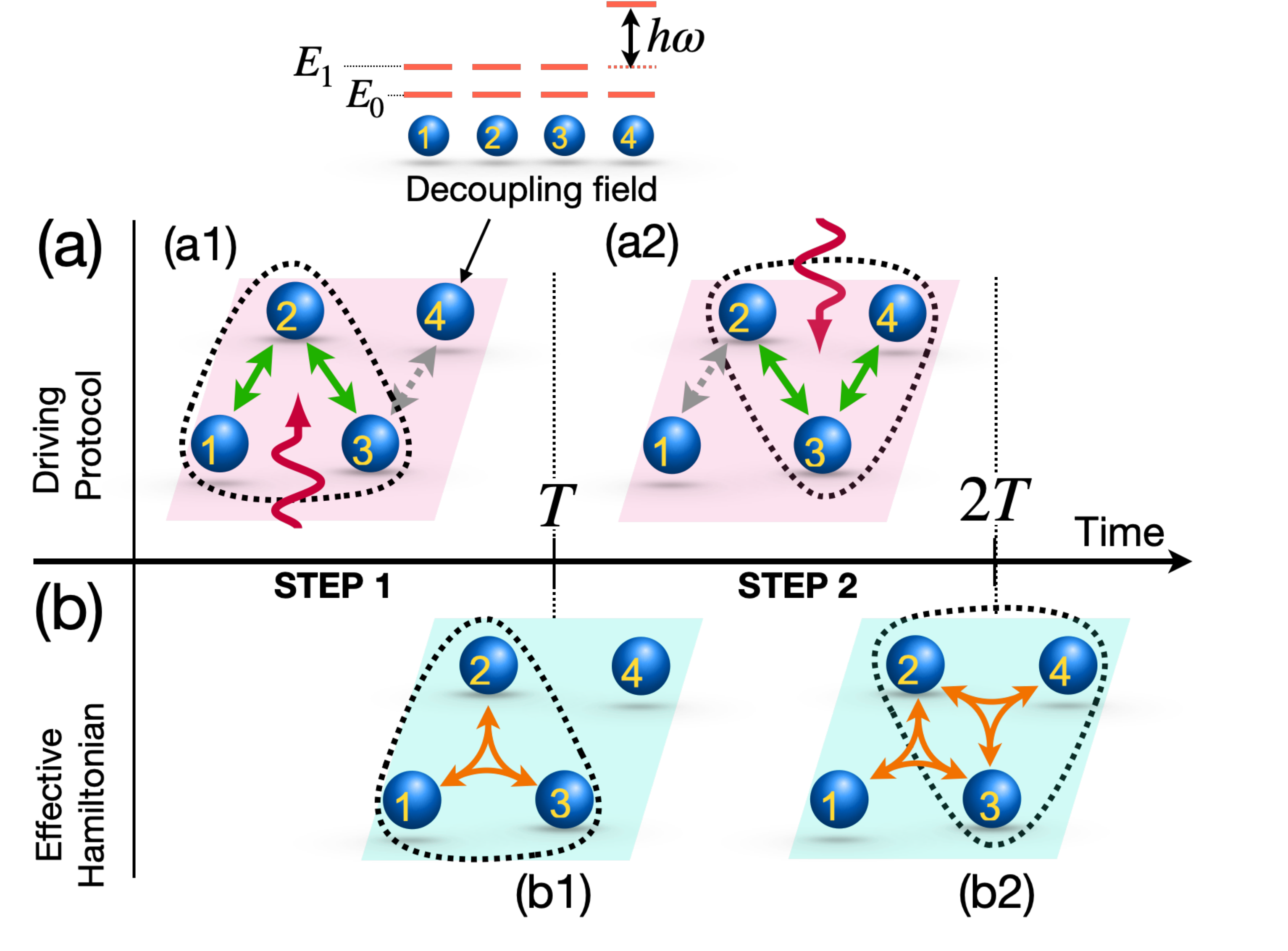}
\caption{Schematics of the digital sequence to simulate the Hamiltonian \eqref{eq:Htg1} on four spins.}
\label{fig:decoupling}
\end{figure}

This section discusses the realization of a pure three-body interaction in a system with non-commuting terms in the Hamiltonian.
The digital part of the quantum simulation thus results in Trotterization errors, and different subsystems need to be decoupled via control fields in the analog part.
The model considered is given by the two-body $XY$ Hamiltonian
\begin{equation}\label{eq:modelXY}
H = J\sum_{k=1}^{N-1} [ \hat{X}_k \hat{X}_{k+1} + \hat{Y}_k \hat{Y}_{k+1} ] = \frac{J}{2}\sum_{k=1}^{N-1} [ \hat{\sigma}_k^+ \hat{\sigma}^-_{k+1} + \hat{\sigma}^-_k \hat{\sigma}^+_{k+1} ] ,
\end{equation}
where $\hat{\sigma}_k^{\pm}  = [\hat{X}_k \pm i \hat{Y}_k]/2$.
 The target three-body Hamiltonian is
 \begin{equation}\label{eq:Htg1}
 H_{\text{tg}} = J_{\text{eff}} \sum_{k=2}^{N-1}[\hat{X}_{k-1} \hat{Z}_k \hat{X}_{k+1} + \hat{Y}_{k-1} \hat{Z}_k \hat{Y}_{k+1}] = \frac{J_{\text{eff}}}{2} \sum_{k=2}^{N-1}[\hat{\sigma}_{k-1}^+ \hat{Z}_k \hat{\sigma}_{k+1}^- + \hat{\sigma}_{k-1}^- \hat{Z}_k \hat{\sigma}_{k+1}^+].
 \end{equation}
This interaction describes flip-flop of excitations between next-to-nearest-neighbours spins with effective tunnelling coefficients conditioned on the state of the in-between spin that result in frustration.

We treat in the following the case of a four-spin chain with open boundary conditions.
For Floquet engineering the target three-body term on a three-spin subsystem --- involving for instance spins (1, 2, 3) ---  it is first necessary to decouple such a subsystem from the rest of the chain. This can be done by applying a static field of the form $\hh \hat{Z}_4$ to the neighbouring spin 4. In this way, the interaction between 3 and 4 becomes off-resonant and spins (1, 2, 3) become effectively uncoupled from spin 4, according to the same principles discussed in Sec. \ref{sec:crosstalk} in the context of spin crosstalk. This is illustrated in Figure \ref{fig:decoupling}(a1). The field $\hh \hat{Z}_4$ of course also produces a single-qubit evolution for spin 4. To avoid this evolution to give undesired single-qubit contribution to the effective Hamiltonian, it is sufficient to choose the strength $\hh$ to be an integer multiple of $\omega$, such that $e^{-i \hh \hat{Z}_4 T} = \mathbb{1}$. 

Considering now a single three-spin subsystem labeled by indexes (1, 2, 3), the target three-body interaction can be engineered by driving all three spins: spins 1 and 3 are driven with a static component each, denoted by $f_{1,0}$ and $f_{3,0}$, respectively. The central spin 2 is driven with a control function of the same form as that used in the example discussed in the main text, namely $f_2(t) = f_{2,0} + f_{2,1} \cos(\omega t) + f_{2,2} \cos(2 \omega t)$. Overall, the driving Hamiltonian is
\begin{equation}
H_{\text{d}}(t) = f_{1,0} \hat{Z}_1 + [ f_{2,0} + f_{2,1} \cos(\omega t) + f_{2,2} \cos(2 \omega t)] \hat{Z}_2 + f_{3,0} \hat{Z}_3 + h \hat{Z}_4.
\end{equation}

For $\omega = 5J$, a set of parameters found via numerical optimisation for a three-spin subsystem is
\begin{align}\label{eq:driveparams}
& f_{1,0} = 0.05553\omega \ , \quad f_{3,0} = 0.05553\omega \ , \nonumber \\
& f_{2,0} = 0.88894 \omega \ , \quad f_{2,1}= 0.75227\omega \ , \quad f_{2,2} = 0.61233\omega.
\end{align}

To achieve the full target Hamiltonian of Eq. \eqref{eq:Htg1}, two digital steps are sufficient, see Fig. \ref{fig:decoupling}. In the first step, the three-body interaction is engineered on spins (1, 2, 3) while spin 4 is driven in order to decouple the subsystem (1, 2, 3) [Fig. \ref{fig:decoupling}(a1) and (b1)]. In the second step, spin 1 is decoupled while the three-body term is engineered on (2, 3, 4) [Fig. \ref{fig:decoupling}(a2)]. 
At the end of the sequence [Fig. \ref{fig:decoupling}(b2)], the target three-body Hamiltonian \eqref{eq:Htg1} is achieved on the whole system up to digitisation errors of order $O(T^2)$ and decoupling errors.
\begin{figure}
\includegraphics[width=\textwidth]{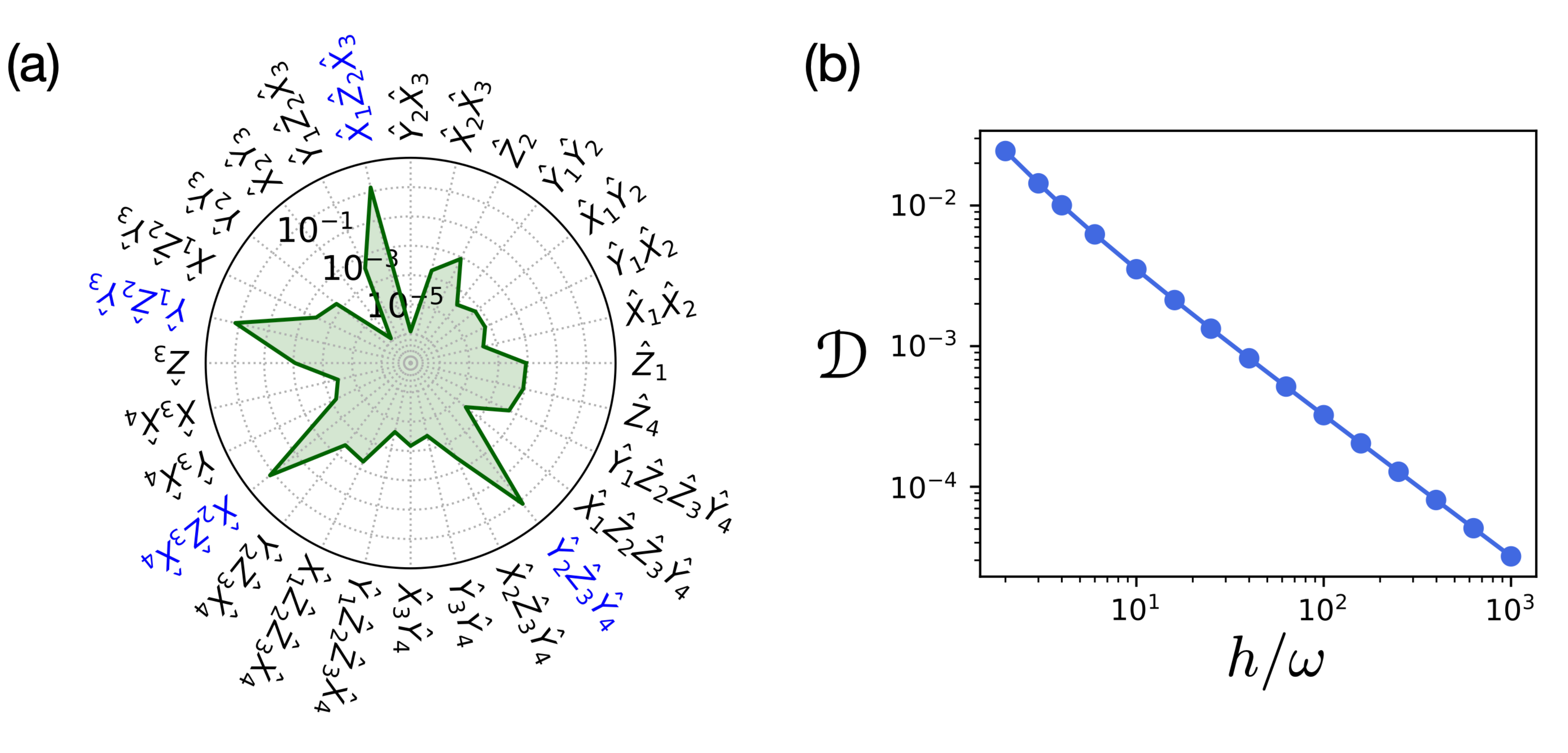}
\caption{(a) Chart of the absolute value of the components of the effective Hamiltonian at the end of the digital sequence, using $\omega=5J$, the driving parameters of Eq. \eqref{eq:driveparams} and $\hh=100\omega$. The effective Hamiltonian is derived from diagonalization of the propagator $U(2T)$. (b) Error induced by the decoupling procedure of spin 4 via control fields in the first step of the digital sequence as a function of the strength $h$ of the decoupling field, for $\omega=5J$ and driving parameters of Eq. \eqref{eq:driveparams}.
}
\label{fig:xymodel}
\end{figure}

The results of the full three-body target Hamiltonian on the four-spin chain are reported in Fig. \ref{fig:xymodel}(a), using the driving parameters of Eq. \eqref{eq:driveparams} and other relevant parameters given in the caption. The desired three-body terms are successfully produced and are stronger by more than two orders of magnitude as compared to undesired interactions, while still using a relatively weak driving frequency.

In order to characterize errors produced by the decoupling procedure of subsystems,
Fig.~\ref{fig:xymodel}(b) depicts the relative distance $\mathcal{D}$ of Eq. \eqref{eq:norms} between
\begin{itemize}
\item[(i)] the ‘‘ideal'' effective Hamiltonian $\he^{(0)}(T)$ for the four-spin chain at the end of the first digital step assuming that spin 4 is not interacting with spins (1, 2, 3) [\emph{i.e.}, the term $\hat{X}_3 \hat{X}_4+ \hat{Y}_3 \hat{Y}_4$ is removed from the Hamiltonian of Eq. \eqref{eq:modelXY}],
\item[(ii)]  the effective Hamiltonian $\he(T)$ at the end of the first digital step if spin 4 is decoupled via the control field from spins (1, 2, 3) [\emph{i.e.}, the decoupling field $\hh \hat{Z}_4$ is included in the driving Hamiltonian],
\end{itemize}
as a function of the strength $\hh$ of the decoupling field. For a decoupling field larger than a few units of $\omega$, the relative difference of the two effective Hamiltonians is smaller than 1\%.

\end{document}